\documentclass[journal=jacsat,manuscript=article]{achemso}

\usepackage[utf8]{inputenc}
\usepackage{xspace}
\usepackage{dcolumn}
\usepackage{textcomp}
\usepackage{bm}
\usepackage{geometry}
\usepackage{float,subcaption}
\usepackage{tikz}
\usepackage{booktabs,multirow}
\usetikzlibrary{positioning}
\usepackage{amsmath,amssymb}
\usepackage{graphicx}
\usepackage{xcolor}
\DeclareUnicodeCharacter{0301}{\'{e}}

\title{Gas–solid Reaction Dynamics on Li$_6$PS$_5$Cl Surfaces: A Case Study of the Influence of CO$_2$ and CO$_2$/O$_2$ Atmospheres Using AIMD and MLFF Simulations}

\author{Zicun Li}
 \affiliation{College of Physics
Nanjing University of Aeronautics and Astronautics (NUAA)
Nanjing 211106, China}
 \alsoaffiliation{Beijing National Laboratory for Condensed Matter Physics, Institute of Physics, Chinese Academy of Sciences, Beijing 100190, China}
\author{Xinguo Ren}
\email{renxg@iphy.ac.cn}
\affiliation{Beijing National Laboratory for Condensed Matter Physics, Institute of Physics, Chinese Academy of Sciences, Beijing 100190, China}
\author{Jinbin Li}
\email{jinbin@nuaa.edu.cn}
\affiliation{College of Physics
Nanjing University of Aeronautics and Astronautics (NUAA)
Nanjing 211106, China}
\author{Ruijuan Xiao}
\email{rjxiao@iphy.ac.cn}
\affiliation{Beijing National Laboratory for Condensed Matter Physics, Institute of Physics, Chinese Academy of Sciences, Beijing 100190, China}
\author{Hong Li}
\affiliation{Beijing National Laboratory for Condensed Matter Physics, Institute of Physics, Chinese Academy of Sciences, Beijing 100190, China}

\begin{document}

\begin{abstract}
In recent years, rapid progress has been made in solid-state lithium batteries. Among various technologies, coating the surface of electrodes or electrolytes has proven to be an effective method to enhance interfacial stability and improve battery cycling performance. Recent experimental studies showed that gas-solid reactions offer a convenient approach to form modified coating layers on the solid electrolyte. Here, we performed computational simulations to investigate this surface reaction process. Specifically, we simulated the gas-solid reactions of Li$_6$PS$_5$Cl(LPSC) solid-state electrolytes in pure CO$_2$ and in mixed CO$_2$/O$_2$ atmospheres using ab-initio molecular dynamics (AIMD) and machine-learning force fields (MLFF)-accelerated molecular dynamics (MD) approaches. In the former case, LPSC surfaces primarily form Li$_2$CO$_2$S because it is difficult to dissociate another oxygen atom from the second CO$_2$ molecule. While in CO$_2$/O$_2$ mixed atmosphere, O$_2$ molecules preferentially adsorb onto LPSC, which supplies oxygen sites for subsequent CO$_2$ adsorption to form carbonate -CO$_3$ units. This reaction pathway ultimately generates an interfacial product dominated by Li$_2$CO$_3$. These coatings exhibit distinct electronic and ionic conductivity characteristics, allowing the possibility to control coating compositions and configurations by adjusting the gas-solid reactions. Key criteria for applying this strategy are extracted from the current research.

\end{abstract}

\maketitle

\section{Introduction}
In the field of rechargeable batteries, solid-state batteries(SSBs) that use solid electrolytes are considered  promising for next-generation battery technology due to their unique merits in integration and their potential to solve the longstanding dilemma between high energy density and safety.\cite{liang2022challenges,hatzell2020challenges} Researchers have discovered an increasing number of solid electrolytes with ionic conductivities comparable to those of liquid electrolytes, such as Li argyodites\cite{zhao2021excellent,adeli2019boosting,zhou2019new}, Li$_{10}$GeP$_2$S$_{12}$ and derivatives\cite{kamaya2011lithium,kato2016high}, garnet structure of Li$_x$La$_3$M$_2$O$_{12}$(5$\le$x$\le$7, M = Nb, Ta, Sb, Zr, Sn)\cite{miwa2017interatomic,bernuy2014atmosphere}, moisture-stable Li$_2$BMS$_4$(B = Ba, Sr; M = Sn, Si)\cite{mao2024moisture}, pyrochlore-type Li$_{2-x}$La$_{(1+x)/3}$Nb$_2$O$_6$F system\cite{aimi2024high}, etc., making solid-state lithium batteries increasingly feasible in practical applications. However, one of the major challenges for SSBs lies at the interface between the electrode and the electrolyte. During the charging and discharging processes, the occurrence of side reactions at the interfaces can lead to higher interfacial resistance, while dynamic changes in interfacial stress may cause cracks at the interface, further increasing the interfacial resistance. These factors can adversely affect the battery's cycle life and rate performance.\cite{tan2019elucidating,wang2021interfacial,zhang2020unraveling,richards2016interface,wenzel2016direct} 

\noindent For sulfide solid electrolytes, the recently developed new method involves constructing 
a Li$_2$CO$_3$ interface by treating the electrolyte Li$_6$PS$_5$Cl (LPSC) in CO$_2$-rich atmosphere through a spontaneous gas–solid reaction.\cite{zhang2023spontaneous} Instead of cathode coating approaches, such simple strategy can modify the electrolyte surface uniformly by in-situ formation of a carbonate layer to effectively enhance the moisture toleration and interfacial stability of the sulfide electrolytes. Combined with bare LiCoO$_2$ electrode, the SSBs with CO$_2$/O$_2$-treated Li$_6$PS$_5$Cl can alleviate the side reactions, leading to lower polarization, significantly improved reversible capacity and Coulombic efficiency. This offers a straightforward approach to improve the battery performance. However, it requires in-depth understanding of the reaction processes at the atomic level to design such spontaneous gas-phase reactions effectively. Through the experiment, X-ray photoelectron spectroscopy (XPS) and Fourier transformed infrared spectra (FTIR) detected the increased carbonate (CO$_3^{2-}$) signal as the exposure time extended for LPSC treated in CO$_2$ or CO$_2$/O$_2$ environment. The detailed atomistic mechanism of the gas-solid reaction at the LPSC surface is difficult to directly characterize in experiments, and the intermediate and final products of the reaction cannot be determined directly. Here we conduct computational simulations to study the process of above reactions, trying to offer insights from an alternative research method to understand this modification strategy. To describe the reaction process comprehensively, it is necessary to construct a suitable surface structure and conduct long-term simulations. However, although traditional density functional theory(DFT) provides high accuracy, it remains limited in efficiency when dealing with large surface/interface systems. Regarding processes involving kinetics, current simulation method, such as ab-initio molecular dynamics (AIMD), is limited to a few hundred picoseconds, which is inadequate for describing the complex and dynamically evolving surface reactions. In recent years, the rise of neural networks has led to the development of various machine-learning force fields (MLFF) that ensure accuracy and efficiency, making previously daunting calculations feasible.\cite{miwa2017interatomic,zhang2018deep,bartok2010gaussian,li2025mechanistic}

\noindent In this work, we conduct first-principles calculations combined with MLFFs, to examine the formation and breaking of bonds between CO$_2$/O$_2$ molecules and the LPSC surface, through both AIMD and MLFF-based MD simulations. Based on the calculations, we investigated the atomic structures of the intermediate products and the pathways leading to the possible final products. Additionally, we made computation-based predictions of battery-relevant properties of the products formed on the surface, including the ionic and the electronic conductivities. From these studies, potential design strategies were extracted, offering guidelines for applying this method to design reaction processes for various electrolyte systems.

\section{Methods}
We conducted simulations on the gas-solid reaction at LPSC surface in CO$_2$ and CO$_2$/O$_2$ atmospheres through first-principles calculations and MLFF. Firstly, the adsorptions of a single CO$_2$ molecule on LPSC (100), (110), and (111) surfaces were studied to evaluate the energy and configuration for this event. Considering the lower adsorption energy of LPSC(100), we constructed the model for this stable plane with symmetrical upper and lower surfaces. In this model, there are six PS$_4$ layers in total and a vacuum region with the thickness of 20~\AA. The atomic stoichiometry is maintained in the system. Secondly, we introduced two kinds of gas environment on LPSC (100) surface model, one is adding 10 CO$_2$ molecules and the other is adding 10 CO$_2$ and 4 O$_2$ molecules within the vacuum region to simulate the situation in which the gas molecules are close to the LPSC surface(as shown in Fig.~\ref{aimd_LPSC_CO2}a and Fig.~\ref{LPSC_CO2_O2}a), respectively. The AIMD simulations are performed to find out the possible intermediate configurations and the nudged-elastic-band(NEB) calculations\cite{henkelman2000climbing} are carried out to evaluate the energy landscape on bond-breaking and formation at the surfaces. Following this, the MLFF was trained. In terms of MLFF, we can extend the MD simulations into nanosecond and this enables us to monitor how Li$_2$CO$_2$S or Li$_2$CO$_3$ are formed on the LPSC surface step by step. Finally, the properties of possible products were evaluated through ab-initio calculations to provide understanding on how the coating improve the performance of the battery.

\noindent The first-principles software, the Atomic-orbital Based Ab-initio Computation at UStc (ABACUS), is used for DFT calculations. Specifically, the Perdew-Burke-Ernzerhof (PBE)\cite{perdew1996generalized} generalized gradient approximation is adopted for the exchange-correlation functional and the linear combination of numerical atomic orbital (NAO) basis sets at the level of double-zeta plus polarization (DZP) as developed in Ref~\cite{Lin2021PRB} are used in all PBE calculations. Furthermore, the multi-projector ``SG15-ONCV"-type norm-conserving pseudopotentials are employed to describe the interactions between nuclear ions and valence electrons. The density threshold for the convergence of self-consistent iterations is 10$^{-8}$. A $\Gamma$-only {\bf k}-point is used for the AIMD simulations and $4\times 4 \times 4$ {\bf k}-mesh is used for
the electronic structure calculations. Previously, ABACUS with such settings has been successfully used to study battery material systems\cite{ABACUS_graphite,Wang/etal:CPB:2021,li2025mechanistic}. 
Based on the DFT simulations, the training datasets are constructed. The NequIP\cite{batzner20223} software is used for training the MLFF for the six-element system, including Li, P, S, Cl, O, and C. For training the force-fields by NequIP, we set cutoff radius as 8.0~\AA, the number of interaction blocks as 4 and the batch size as 5. After 1060 epochs, the mean absolute errors(MAE) of energy and force are reduced to 0.5~meV/atom and 9.5~meV/\AA, respectively. The obtained MLFF is employed in the following MD simulations by LAMMPS code\cite{plimpton1995fast}. For MD calculations by LAMMPS, the NvT ensembles are used at the timestep of 0.001~ps to simulate the reaction process in the timescale of nanoseconds. For the specific parameters, the `units' is selected as `metal', atom style is used as `atomic', and the neighbor is set to 3.0 bin. We also determined the Li$^+$ diffusion coefficients at various temperatures to assess the ion conductivity of the system.

\section{Results and Discussion}
\subsection{Adsorption Energies and Structures}
\begin{figure*}
    \centering
    \includegraphics[width=1.0\columnwidth]{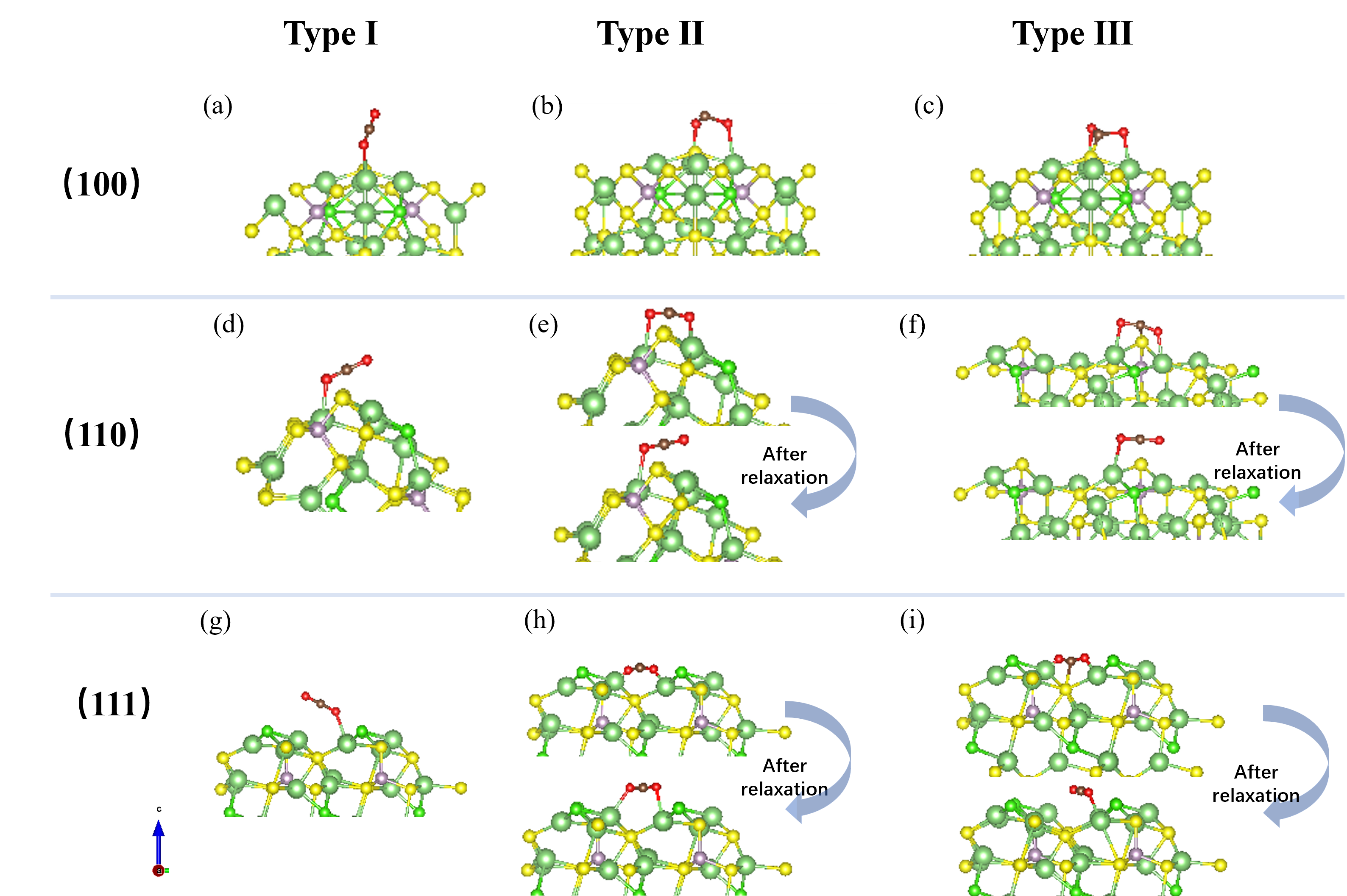}
    \caption{The initial and relaxed configurations for single CO$_2$ molecule adsorbed on LPSC surfaces. Three types of the final CO$_2$ adsorption sites on (a-c) LPSC(100) surface,
    (d-f) LPSC(110) surface and (g-i) LPSC(111) surface. For cases where the configuration type changes after optimization, the initial structure is also provided. The colors of Li, P, S, Cl, O, and C are green, purple, yellow, fluorescent green, red, and brown respectively. }
    \label{adsorption}
\end{figure*}

The previous investigations have reported that the charge transfer can occur from the surface to the adsorbed CO$_2$ molecule, leading to the formation of bended CO$_2^{\gamma -}$\cite{burghaus2014surface,solymosi1991bonding,wang2002adsorption,ernst1999adsorption}, which is considered as precursors to carbonate formation or dissociative adsorption of CO$_2$. Therefore, we simulated various possible configurations of one CO$_2$ molecule on the LPSC surface and carried out structural relaxation on these configurations. Three initial adsorption configurations are considered: the first case is that one oxygen atom of CO$_2$ molecule approaches Li$^{+}$ on the LPSC surface (Fig.~\ref{adsorption} type \uppercase\expandafter{\romannumeral1}). The second case is that two oxygen atoms are connected to two Li$^{+}$ ions without the bond formation between C atom of CO$_2$ and S atom on the LPSC surface(Fig.~\ref{adsorption} type \uppercase\expandafter{\romannumeral2}); the third case is similar to the second one but with the formation of C-S bond(Fig.~\ref{adsorption} type \uppercase\expandafter{\romannumeral3}). The adsorption energy can be calculated using the following equation:
$$
E_{ad}=E_{tot} - E_{CO_2} - E_{surf}
$$
$E_{tot}$, $E_{CO_2}$ and $E_{surf}$ are the energies after relaxation for adsorption configuration, single CO$_2$ molecule and free LPSC surface, respectively. All the adsorption models shown in Fig.~\ref{adsorption} are final configurations after relaxation. The thickness of the slab model, the number of Li$^+$ layers and the adsorption energy $E_{ad}$ for each type of surface with varied Miller indices are listed in Table.~\ref{Adsorption_Energy_co2}. The adsorption energies of type \uppercase\expandafter{\romannumeral1} on different surfaces are around -0.1~eV, and the adsorption energy with small absolute value indicates that it is more likely to form physical adsorption with this configuration. The adsorption energy of type \uppercase\expandafter{\romannumeral2} on the (100) surface is positive indicating this is an unstable configuration. The initial structures of adsorption on the (110) and (111) surfaces after structural relaxation evolve to type \uppercase\expandafter{\romannumeral1}. Adsorption type \uppercase\expandafter{\romannumeral3} shows the strongest adsorption energy on the (100) surface, but the relaxation for adsorption structures on the (110) and (111) surfaces return to adsorption type \uppercase\expandafter{\romannumeral1}. The configuration in Fig.~\ref{adsorption}c is the most stable one, because charge transfer occurs between CO$_2$ and the exposed S atom on the LPSC(100) surface, forming stable C-S bond, which was also detected during CO$_2$ adsorption through XAS spectra in experimental observation\cite{jiang2023enhancing}. Therefore, the LPSC(100) surface is chosen in the subsequent simulations to study the evolution of the gas-solid interface between CO$_2$ and LPSC.

\begin{table}[]
    \centering
    \begin{tabular}{p{2cm}<{\centering}p{3cm}<{\centering}p{3cm}<{\centering}p{2cm}<{\centering}p{3cm}<{\centering}p{2cm}<{\centering}}
      \toprule
      Miller indices (hkl) & thickness of the slab (\AA) & number of Li$^+$ layers & model index corresponding to Fig.~\ref{adsorption} &Adsorption Sites &  $E_{ad}$ (eV) \\
      \midrule
        {\multirow{2}{*}{(100)}} & {\multirow{2}{*}{30.84}} & {\multirow{2}{*}{13}} & a&Type \uppercase\expandafter{\romannumeral1} & -0.05 \\
        {\multirow{2}{*}{}}&{\multirow{2}{*}{}} &{\multirow{2}{*}{}} & b& Type \uppercase\expandafter{\romannumeral2} & 0.17 \\
        {\multirow{2}{*}{}}&{\multirow{2}{*}{}} &{\multirow{2}{*}{}} &c& Type \uppercase\expandafter{\romannumeral3} & -0.48 \\
        \midrule
        {\multirow{2}{*}{(110)}}& {\multirow{2}{*}{22.14}} & {\multirow{2}{*}{13}} &d& Type \uppercase\expandafter{\romannumeral1}& -0.14 \\  
        {\multirow{2}{*}{}}&{\multirow{2}{*}{}} &{\multirow{2}{*}{}} & e&Type \uppercase\expandafter{\romannumeral2}&  - \\
        {\multirow{2}{*}{}}&{\multirow{2}{*}{}} &{\multirow{2}{*}{}} & f&Type \uppercase\expandafter{\romannumeral3}&  -  \\    
        \midrule
        {\multirow{2}{*}{(111)}} & {\multirow{2}{*}{21.46}} & {\multirow{2}{*}{8}} & g&Type \uppercase\expandafter{\romannumeral1}& -0.09 \\ 
        {\multirow{2}{*}{}}&{\multirow{2}{*}{}} &{\multirow{2}{*}{}} &h& Type \uppercase\expandafter{\romannumeral2} &  - \\
        {\multirow{2}{*}{}}&{\multirow{2}{*}{}} &{\multirow{2}{*}{}} & i&Type \uppercase\expandafter{\romannumeral3} &  - \\        
     \bottomrule
    \end{tabular}
    \caption{Calculated adsorption energy, $E_{ad}$, of single CO$_2$ molecule on different LPSC surfaces with various initial configuration. The thickness of the slab model, number of Li$^{+}$ layers, and the model index corresponding to Fig.~\ref{adsorption} are also provided for each structure.}
    \label{Adsorption_Energy_co2}
\end{table}

\subsection{Reaction in CO$_2$ Atmosphere}
\begin{figure*}
    \centering
    \includegraphics[width=0.8\columnwidth]{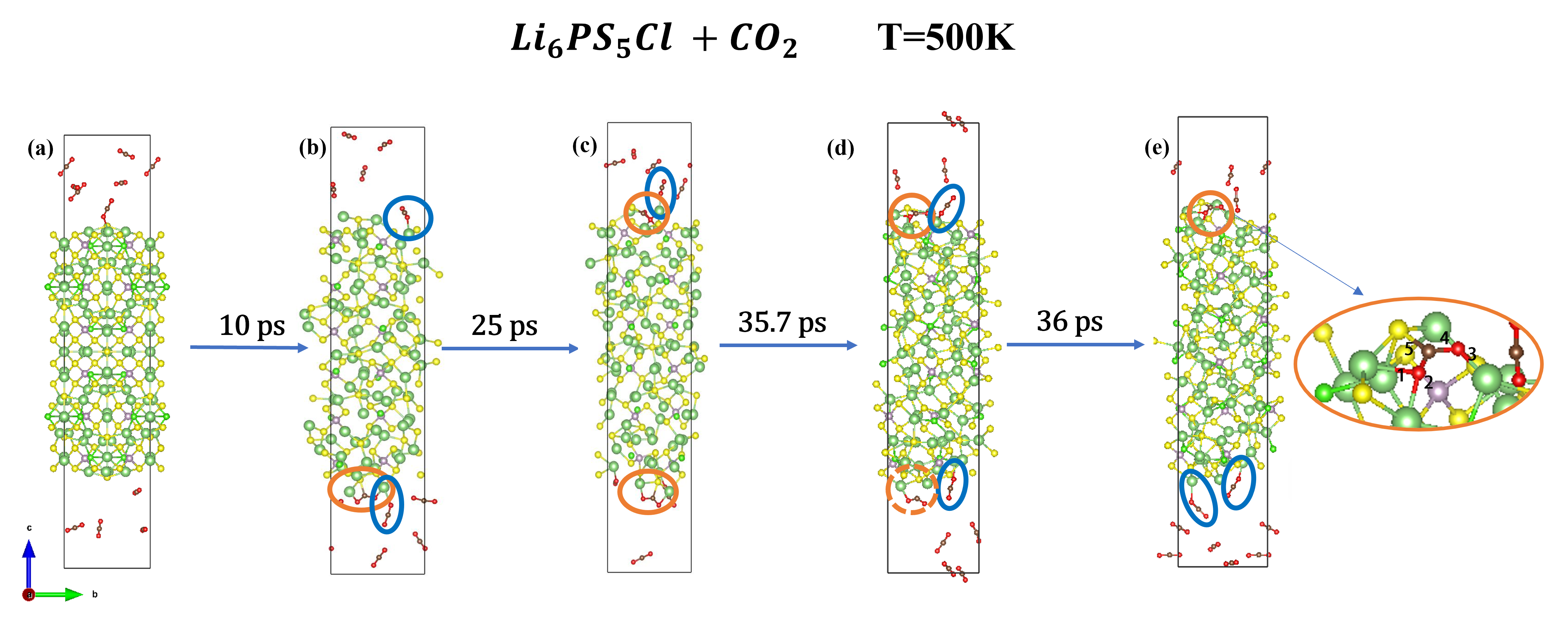}
    \caption{The AIMD simulations of the reaction process on the LPSC(100) surface in a pure CO$_2$ atmosphere is performed using NvT ensembles at 500~K. The system is modeled with ten CO$_2$ molecules, and the configurations at different time intervals: (a) The initial structure, and (b), (c), (d) and (e) are configuration at 10~ps, 25~ps, 35.7~ps and 36~ps respectively; The -CO$_2$S unit in structure (e) is enlarged, clearly illustrating the specific number of bonds between CO$_2$ and LPSC. The blue circles represent the CO$_{2}$ configuration adsorbed on the LPSC(100) via type I mode, and the orange circles correspond to the type III adsorption model which forms the -CO$_{2}$S. And intermediate structures are marked within dashed circles.}
    \label{aimd_LPSC_CO2}
\end{figure*}

The calculated adsorption energy on the LPSC(100) surface indicates that CO$_2$ can spontaneously adsorb onto this surface, forming two Li-O bonds and one C-S bond between CO$_2$ and LPSC. This configuration not only causes the linear CO$_2$ molecule to bend but also forms a planar triangular -CO$_2$S unit similar to carbonate. Based on this, we infer that the bonding and charge transfer processes between CO$_2$ molecules and the LPSC(100) surface can occur rapidly as CO$_2$ approaches the LPSC surface. To observe the formation process of the -CO$_2$S unit, we introduced 10 CO$_2$ molecules into the vacuum region of the LPSC(100) surface model, as illustrated in Fig.~\ref{aimd_LPSC_CO2}. We conducted AIMD simulations with NvT ensembles at the temperature of 500~K. The initial model shown in Fig.~\ref{aimd_LPSC_CO2}a was built in a symmetric manner to ensure that both the top and the bottom surfaces are equivalent. Ten CO$_2$ molecules were then randomly placed within the vacuum layer. This model simulates the LPSC(100) surface exposed to the CO$_2$ gas environment, corresponding to the situation where pure CO$_2$ is introduced in the experiment. Fig.~\ref{aimd_LPSC_CO2} shows the reaction process between LPSC(100) and CO$_2$ within 45~ps at 500~K. In the initial structure, one CO$_2$ connects to the LPSC(100) surface with the type I configuration within the blue circle described in Fig.~\ref{adsorption}, while at 10 ps, two CO$_2$ molecules in type I adsorption model and one CO$_2$ molecule in type III model are formed, as illustrated in Fig.~\ref{aimd_LPSC_CO2}b. As the simulation progressed, type III structures emerged on both the upper and lower surfaces at 25~ps. The C-S bonds within the orange circle below remained stable until 35.7~ps, after which they began to dissociate. At 36~ps, the -CO$_2$S unit reverts to a CO$_2$ molecule, and no further reactions are observed up to 45~ps. This instability is likely attributed to the limited number of bonds formed between CO$_2$ and LPSC surface. One of the two CO$_2$ molecules adsorbed in type I model at 10~ps leaves the surface during the following evolution. This observation confirms that type I is a physical adsorption mode, as inferred from the $E_{ad}$ values. We also performed the AIMD simulations at higher temperatures including 1000~K. On one hand, the use of higher temperatures accelerates the kinetic evolution process on a practically feasible AIMD timescale. On the other hand, simulations at higher temperatures can capture a wider variety of structural configurations, providing a comprehensive dataset for subsequent MLFF training. At elevated temperatures, CO$_2$ consistently reacts with the LPSC(100) surface, producing only -CO$_2$S as a product, with the reaction rate increasing with temperature.

\begin{figure*}[htbp]
    \centering
    \includegraphics[width=1.0\columnwidth]{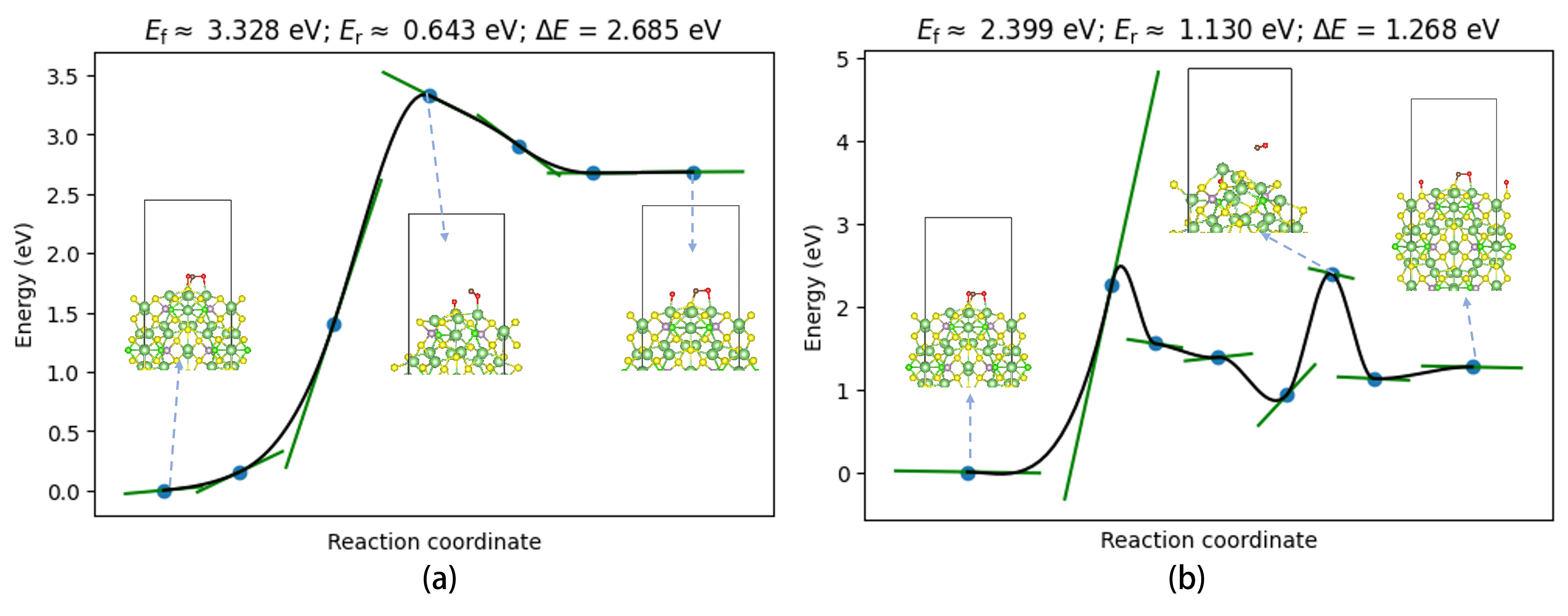}
    \caption{The energy curves corresponding to the bond-breaking of CO$_2$ adsorbed on the LPSC(100) surface calculated by the NEB method for (a) the pathway where the dissociated O binds with Li$^+$ as the final state, and (b) the pathway where the released O binds with S after the C-O bond is broken. $E_f$, $E_r$, and $\Delta E$ represent forward energy barrier, reverse energy barrier, and reaction energy, respectively. }
    \label{broken}
\end{figure*}

\noindent In the above simulations, no further reaction, e.g. forming -CO$_3$ unit, occurred within the limited simulation time of 45~ps. To evolve from a -CO$_2$S structure to a -CO$_3$ unit, it is necessary not only to break the C-S bond but also to obtain an extra oxygen atom from another CO$_2$ molecule, implying that CO$_2$ needs to undergo bond-breaking on the LPSC surface. Thus, we performed the nudged-elastic band(NEB) simulations to evaluate the energy barriers for this process and to estimate whether the reaction could happen. The structure depicted in Fig.~\ref{adsorption}c was selected as the initial configuration. Following the cleavage of the C–O bond in CO$_2$, the dissociated oxygen atom either binds to a Li$^+$ on the LPSC surface, defining Final State illustrated in Fig.~\ref{broken}a, or forms a bond with sulfur (S), corresponding to Final State in Fig.~\ref{broken}b. For the two pathways represented in Fig.~\ref{broken}a and Fig.~\ref{broken}b, the thermodynamic energy differences($\Delta E =E_{fin} -E_{ini}$) for initial and final states are 2.685~eV and 1.268~eV, respectively. And the energy difference between the transition state energy and the initial state energy, denoted as forward energy barrier $E_f$($E_f = E_{ts} - E_{ini}$), as well as the energy difference between the transition state energy and the final state energy, denoted as reverse energy barrier $E_r$($E_r = E_{ts} -E_{fin}$), are provided. The $E_f$ and $E_r$ of reaction process (a) are 3.328~eV and 0.643~eV, respectively, while of reaction process (b) are 2.399~eV and 1.130~eV, respectively. Regardless of the adsorption configurations in final states and the varied reaction pathways, the high energy barriers larger than 2~eV indicate that it is difficult to break bonds in CO$_2$ adsorbed on the LPSC surface at the simulated temperature and pressure. Theoretical simulations show that the bond-breaking energy of CO$_2$ adsorbed on various metal surfaces ranges from 0.65~eV to 1.71~eV.\cite{liu2012co2} According to above information, we can infer that the products for gas-solid reactions occurring at LPSC(100) surface in pure CO$_2$ atmosphere are mainly Li$_2$CO$_2$S.

\begin{figure*}
    \centering
    \includegraphics[width=0.6\columnwidth]{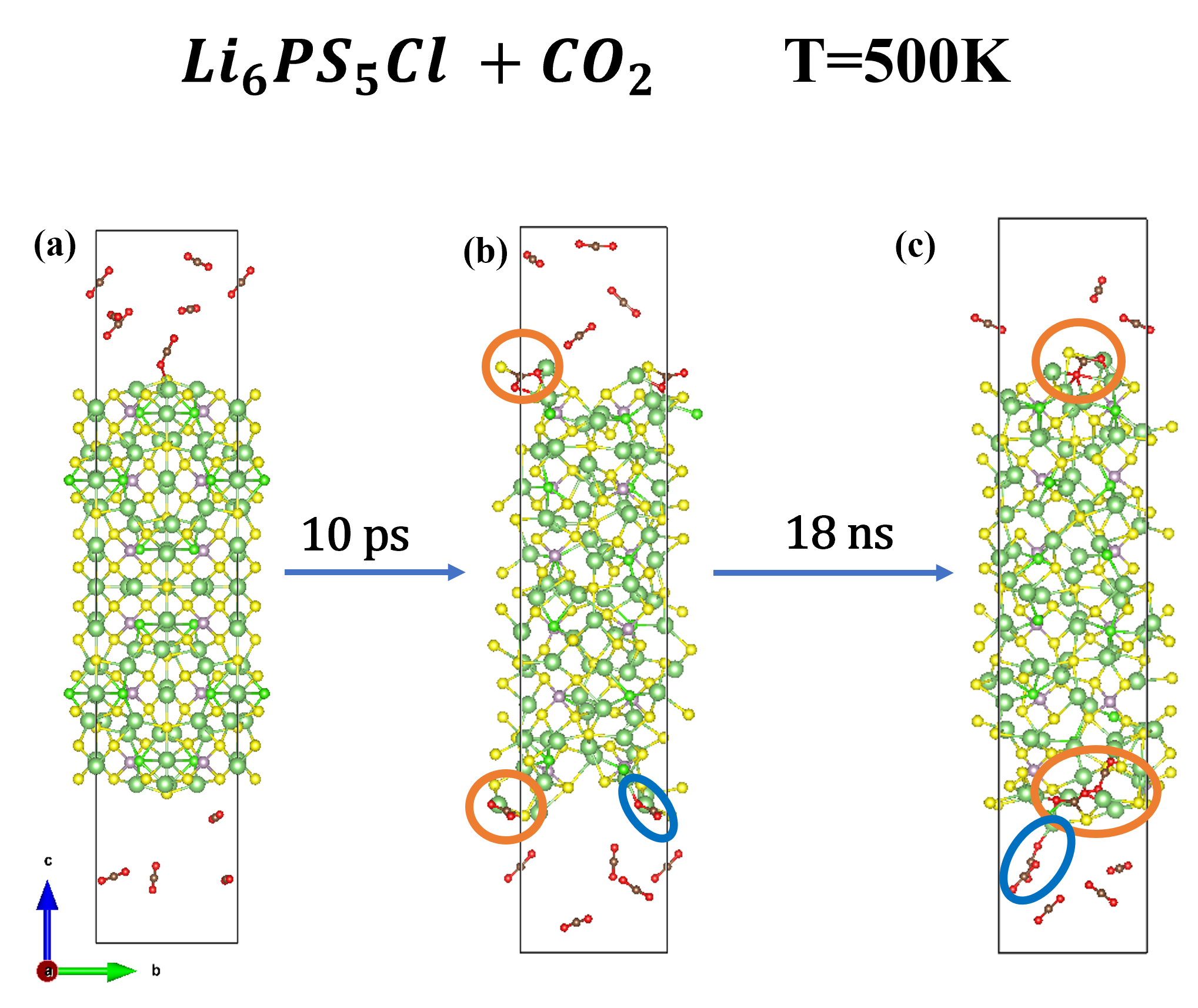}
    \caption{The structure evolution of LPSC(100) surface in pure CO$_2$ atmosphere predicted by MLFF-based MD at 500~K. (a) the initial structure, (b) configuration at 10~ps and  (c) the configuration at 18~ns.}
    \label{mlff_LPSC_CO2}
\end{figure*}

\noindent To explore the evolution of the LPSC(100) surface over longer timescales, we trained an MLFF using the AIMD data obtained above. This MLFF-based MD can significantly exceed both the model size and the timescale of traditional AIMD simulations.\cite{huang2021deep,li2025mechanistic} Therefore, We conducted MLFF-based MD simulations lasting to the nanosecond timescale at different temperatures, 500~K to 900~K. The appearance of -CO$_2$S unit is observed at the early stage within 10~ps as that revealed in the AIMD simulations shown in Fig.~\ref{aimd_LPSC_CO2}b, and an extra -CO$_2$S unit appeared in the orange circle at the lower surface near to 18~ns without forming any other structures during the process. The corresponding structures are shown in Fig.~\ref{mlff_LPSC_CO2}.\\

\begin{figure}
    \centering
    \includegraphics[width=0.7\columnwidth]{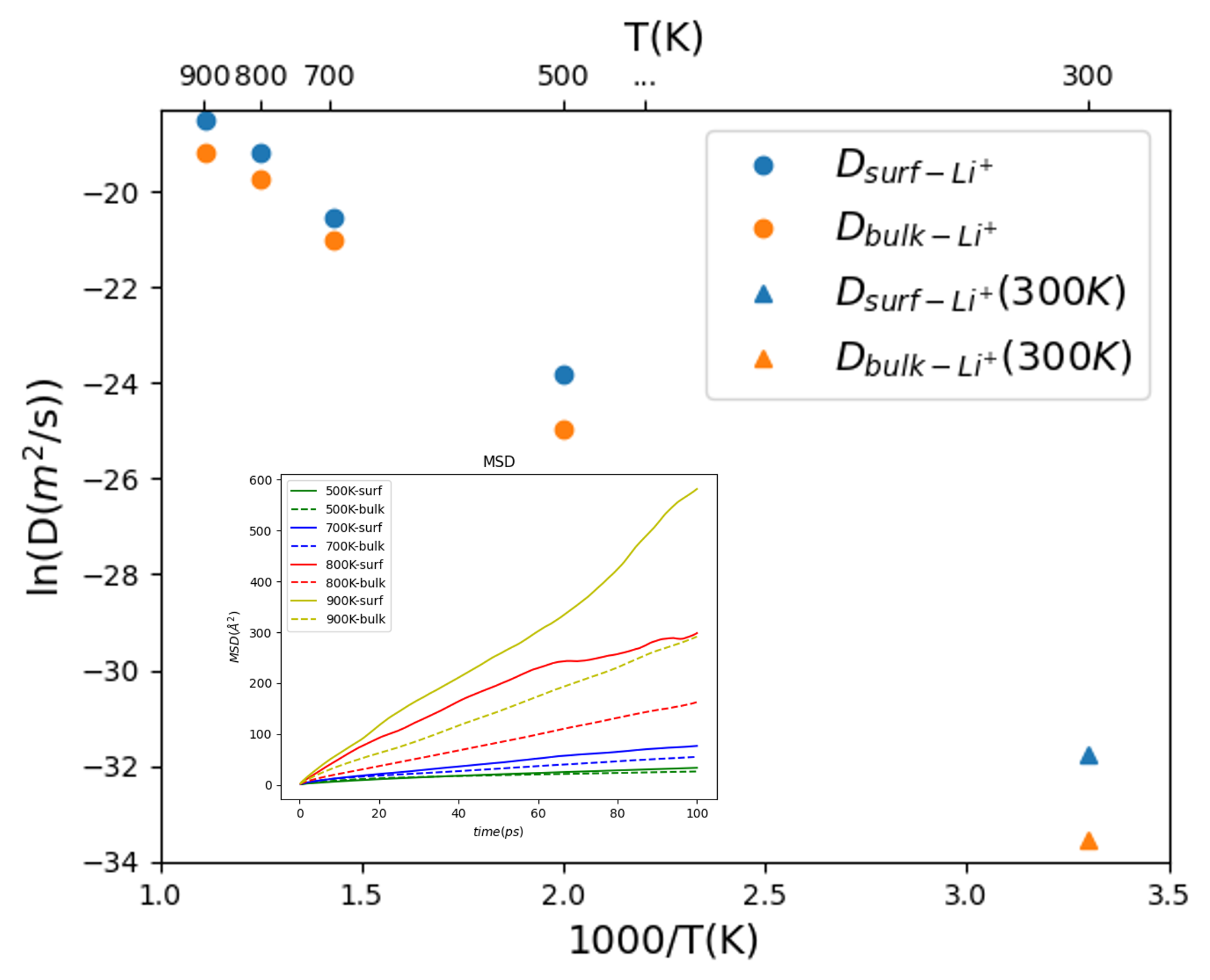}
    \caption{The Li$^{+}$ diffusion coefficients of the surface and bulk regions for the supercell structure at 500~K, 700~K, 800~K, and 900~K simulated with MLFF-based MD. The room temperature (300~K) ionic conductivity is obtained from extrapolation of high-temperature MD results}. 
    \label{fig:ML-lnD-msd}
\end{figure}

\noindent To estimate the effect of gas-solid process reactions on the overall electrolyte conductivity, we carried out the MLFF-based MD simulations in the $2\times2\times1$ supercell structure(as shown in Fig.~S4) for 100~ps at 500~K, 700~K, 800~K, and 900~K. The Li$^{+}$ diffusion coefficients for the surface and the bulk region are determined through Mean Squared Displacement (MSD) data of Li$^+$ by applying the Einstein relation \cite{he2018statistical}, and shown in Fig.~\ref{fig:ML-lnD-msd}. We used the three Li$^+$ layers near to the top and bottom to signify the surface region, and considered the other seven layers close to the center of the 
slab model as the bulk region. At the picosecond scale, it can be seen that the ionic conductivity of the surface part after the reaction is a litter larger than the bulk part, and the higher the temperature, the more obvious it is. Due to the formation of new products and the distinct chemical environments at the surface, the different ionic transport characteristics may appear when compared to the bulk properties.\cite{golov2021molecular,shi2012direct,wang2018review}. Here, the higher ionic conductivity at the surface is mainly caused by the movement of Li$^+$ ions involved in the gas-solid reaction process. The comparison of the Arrhenius relationship between the Li$^+$ ions in bulk and surface regions is given in Fig.~\ref{fig:ML-lnD-msd}, and the extrapolated room-temperature ion conductivities of the surface and bulk regions are $1.27\times 10^{-3}$~S/m and $2.15\times 10^{-4}$~S/m, respectively.

\subsection{Reaction in CO$_2$/O$_2$ Atmosphere}
\begin{figure*}
    \centering
    \includegraphics[width=1.0\columnwidth]{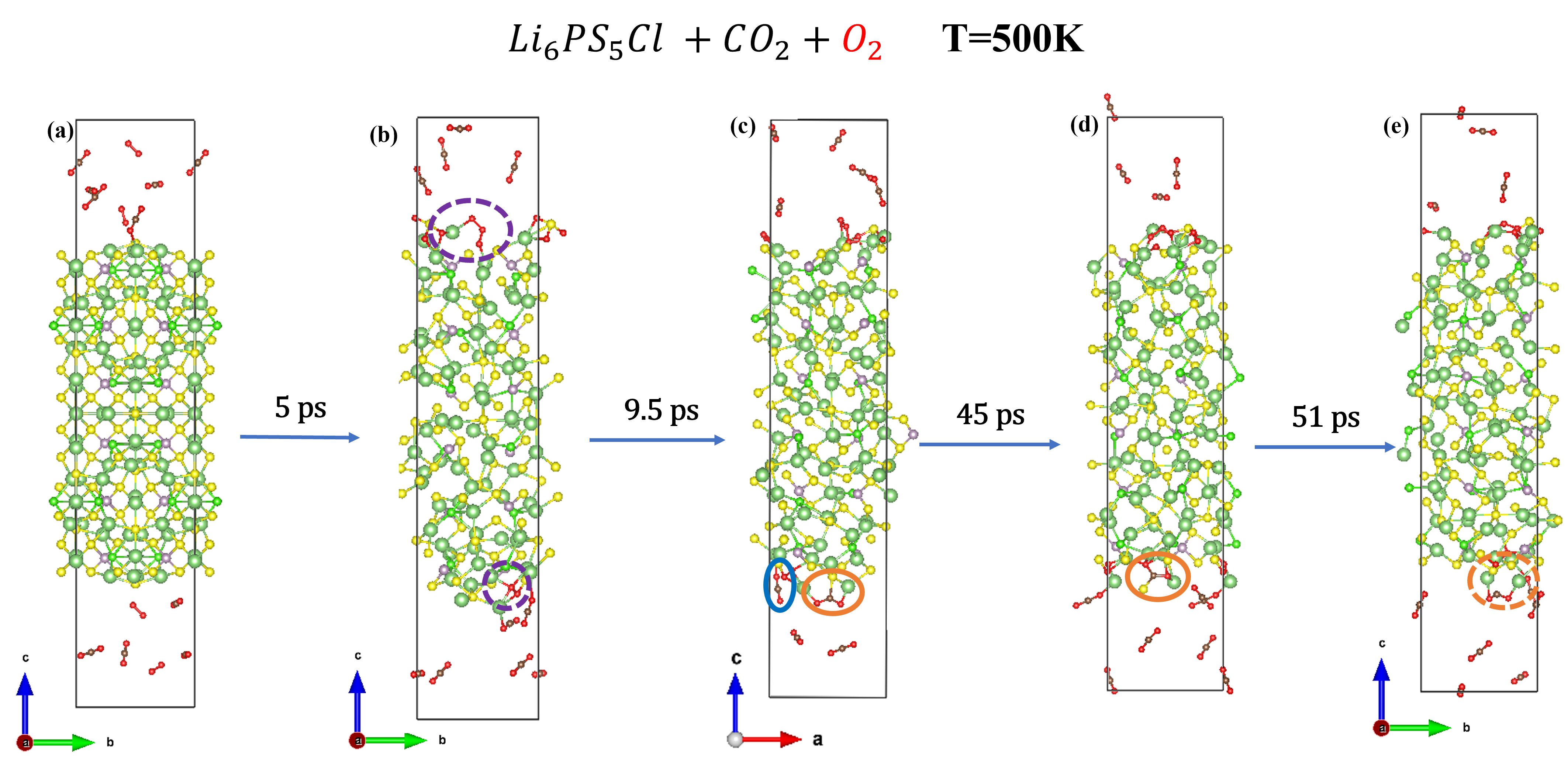}
    \caption{The reaction process of the LPSC(100) surface in a mixed  CO$_2$/O$_2$ atmosphere simulated using NvT-ensemble AIMD at 500~K.}
    \label{LPSC_CO2_O2-500k}
\end{figure*}

\begin{figure*}
    \centering
    \includegraphics[width=1.0\columnwidth]{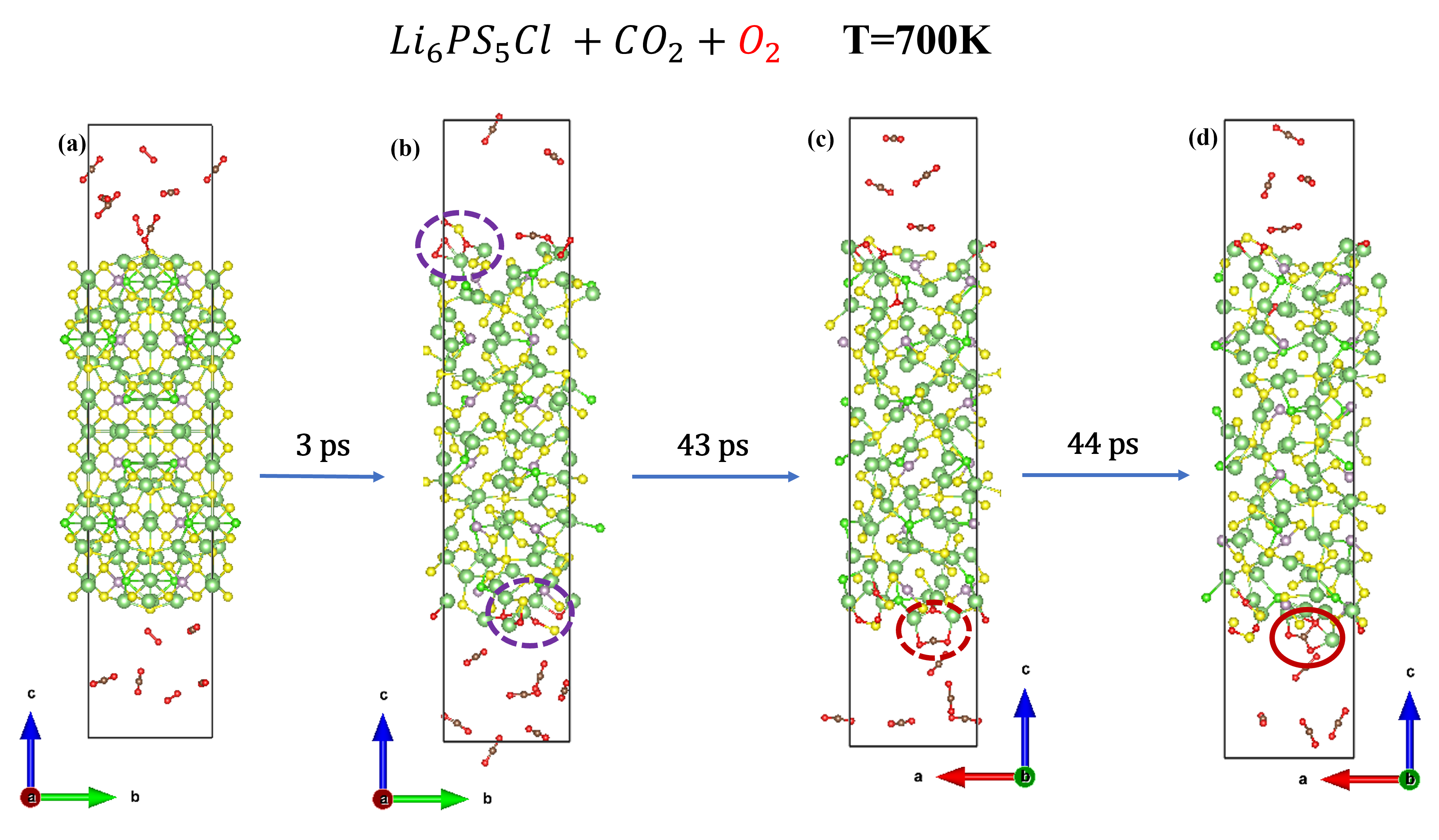}
    \caption{The reaction process of the LPSC(100) surface in a mixed  CO$_2$/O$_2$ atmosphere simulated using NvT-ensemble AIMD at 700~K.}
    \label{LPSC_CO2_O2}
\end{figure*}

According to the MD results for the LPSC(100) surface in CO$_2$ environment, products consisting of Li$_2$CO$_2$S units can form, which are structurally very similar to Li$_2$CO$_3$. However, due to the high energy barriers for C-O bond-breaking, it is difficult to obtain an additional O atom from another CO$_2$ molecule to facilitate the conversion of  Li$_2$CO$_2$S into Li$_2$CO$_3$. Considering that O$_2$ can also be adsorbed on the LPSC surface, thereby providing a source of O atoms, we subsequently simulated the gas-solid reaction processes on the LPSC(100) surface under a CO$_2$/O$_{2}$ mixed atmosphere. The bond-breaking energy of O$_2$ molecules adsorbed on the LPSC(100) surface was calculated, yielding a thermodynamic energy difference ($\Delta E$) of -4.96~eV between the initial and final states (Fig. S5). This negative $\Delta E$ indicates that O$_2$ undergoes spontaneous adsorption and bond-breaking on the LPSC(100) surface, indicating the possibility of Li$_2$CO$_2$S being converted into  Li$_2$CO$_3$ with the assistance of O$_{2}$.  To investigate this process, we constructed a model with the LPSC(100) surface exposed to a mixed CO$_2$/O$_2$ atmosphere. In addition to the ten CO$_2$ molecules, four O$_2$ molecules were introduced into the vacuum region, as illustrated in Fig.~\ref{LPSC_CO2_O2-500k}a. AIMD simulations were performed at 500~K and 700~K for 50~ps to track the reaction dynamics. Within 5~ps, O$_2$ molecule adsorbs onto the LPSC(100) surface, as indicated by the dashed purple circles in Fig.~\ref{LPSC_CO2_O2-500k}b and Fig.~\ref{LPSC_CO2_O2}b. At 500~K, one O$_2$ molecules participate in reactions with the bottom surface of LPSC, whereas the remaining O$_2$ molecules react with the upper surface. The formation of -CO$_2$S unit appeared at 9.5~ps. Similar to the reaction process in a pure CO$_2$ atmosphere, CO$_2$ molecules desorbed from the LPSC surface around 50~ps. At 700~K, O$_2$ molecules reacted uniformly with both the upper and lower surfaces of LPSC at 3~ps, exhibiting a faster reaction rate compared to the 500~K case. As oxygen atoms penetrated the electrolyte surface, they inhibited the adsorption of CO$_2$. By 43~ps, one oxygen atom replaced a free sulfur atom, forming a curved CO$_2^{\gamma -}$ structure shown within the dashed red circle. At 44~ps, -CO$_3$ units emerged by the formation of C-O bond between the  CO$_2^{\gamma -}$ and the O adsorbing on LPSC surface, as shown within the red circle in Fig.~\ref{LPSC_CO2_O2}d. To validate the thermodynamic stability of reaction products, we further calculated the adsorption energy of CO$_2$ on the LPSC surface, where CO$_2$ adsorbs onto the oxygen atom on the surface, as shown in Fig.~S6. The negative adsorption energy value, -2.48~eV, confirms that CO$_2$ adsorbes on the surface oxygen atom, resulting the stable formation of -CO$_3$ unit. By comparing Fig.~\ref{LPSC_CO2_O2-500k} and Fig.~\ref{LPSC_CO2_O2}, we infer that the formation of -CO$_3$ units requires a sufficient O$_2$ concentration in the reaction environment. This suggests that the reaction products can be modulated by adjusting the concentrations of O$_2$ and CO$_2$ molecules. 

\section{Electronic Structure and Conductivity of Lithium Carbonate}
\begin{figure}[!t]
\centering
\subfloat[]{
		\includegraphics[scale=0.5]{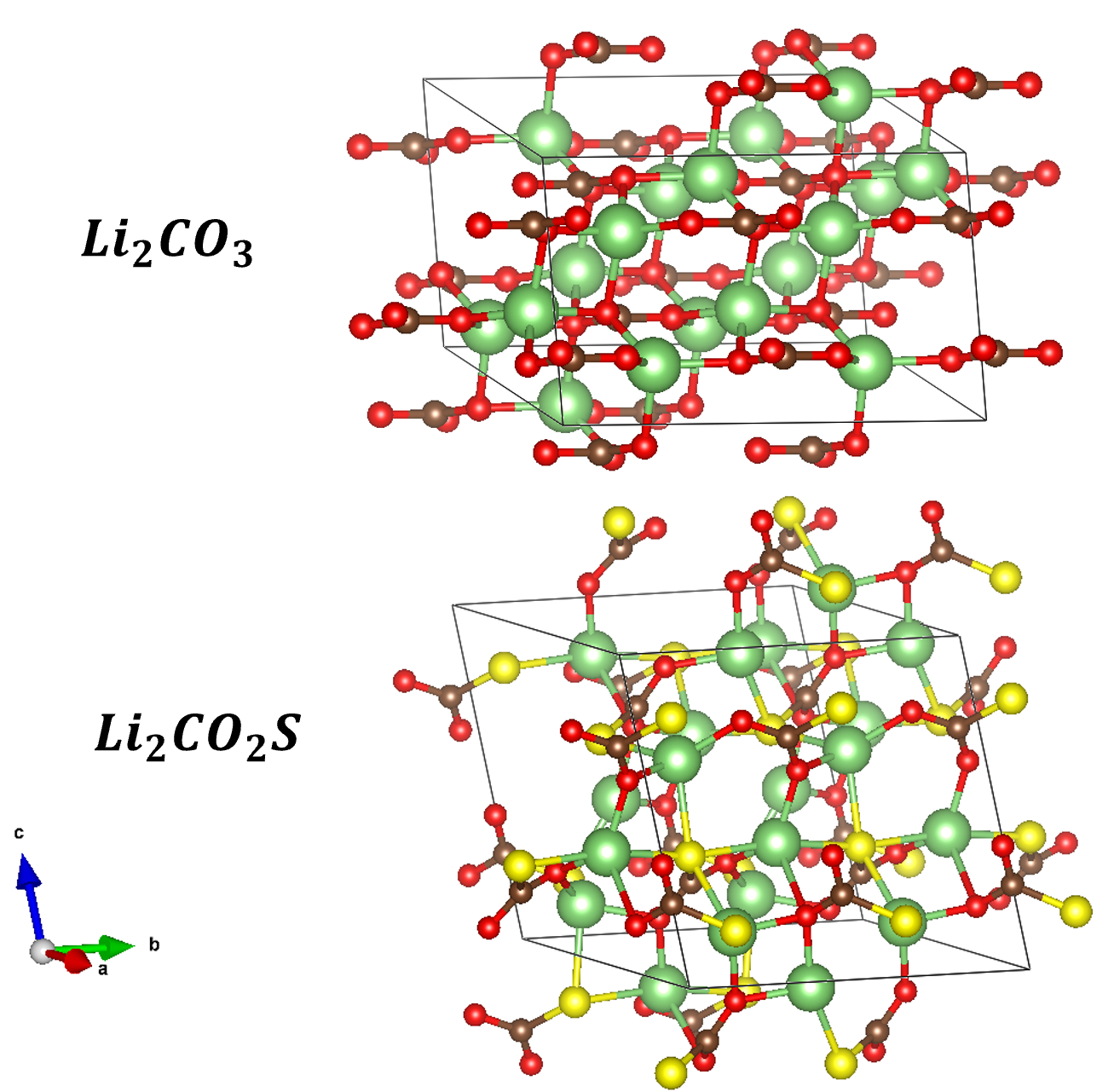}}
\subfloat[]{
		\includegraphics[scale=0.5]{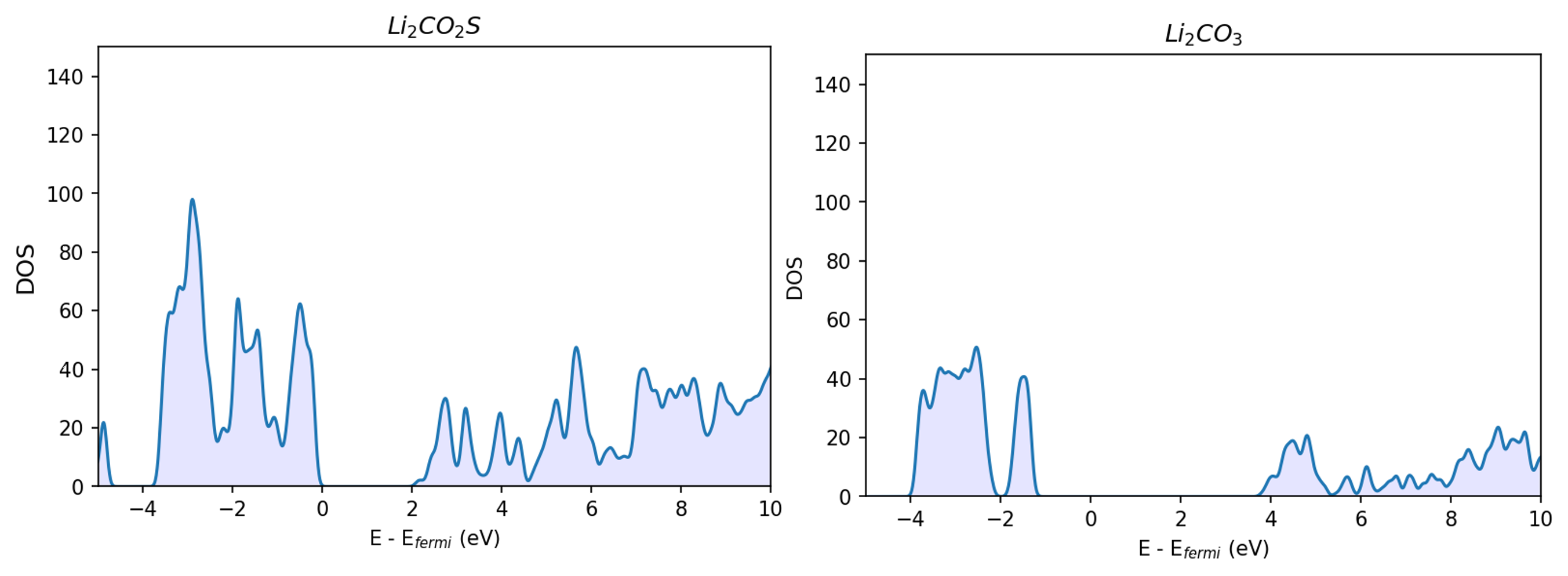}}
\caption{(a)The optimized unit cells of Li$_2$CO$_3$ and Li$_2$CO$_2$S in the monoclinic structure, which are then expanded to form $1 \times 2 \times 1$ supercells, (b)the DOS plot for Li$_2$CO$_3$ and Li$_2$CO$_2$S.}
\label{fig:STRU-DOS}
\end{figure}

Different coating compositions exhibit distinct physicochemical properties, which significantly influence battery performance. To explore these effects, we modeled the crystal systems of Li$_2$CO$_3$\cite{jain2013commentary,ong2015materials} and Li$_2$CO$_2$S in monoclinic structures. The variant Li$_2$CO$_2$S was constructed by replacing one oxygen layer(one-third of oxygen atoms) with a sulfur layer, forming -CO$_2$S units. A $1\times 2\times 1$ supercell was adopted in the simulations, as illustrated in Fig.~\ref{fig:STRU-DOS}. Due to the larger ionic radius of S$^{2-}$ and weaker Li-S bond strength, the optimized Li$_2$CO$_2$S structure exhibits slightly larger lattice parameters compared to Li$_2$CO$_3$.  The lattice parameters of the Li$_2$CO$_3$ are a=8.282~\AA, b=9.9381~\AA, c=6.076~\AA, and $\beta=113.75^{\circ}$, whereas those of Li$_2$CO$_2$S are a=9.035~\AA, b=9.778~\AA, c=8.997~\AA, and $\beta=117.71^{\circ}$. Using the relaxed crystal structure, we evaluated the electronic structure through DFT and ionic conductivity via AIMD of both compositions. The calculated band gap of Li$_2$CO$_3$ is 4.94~eV, consistent with previous reports.\cite{duan2009density,jain2013commentary,ong2015materials} Fig.~\ref{fig:STRU-DOS} indicates that the band gap of Li$_2$CO$_2$S is approximately 2~eV, much smaller than that of Li$_2$CO$_3$. 

\begin{figure*}
    \centering
    \includegraphics[width=1.0\columnwidth]{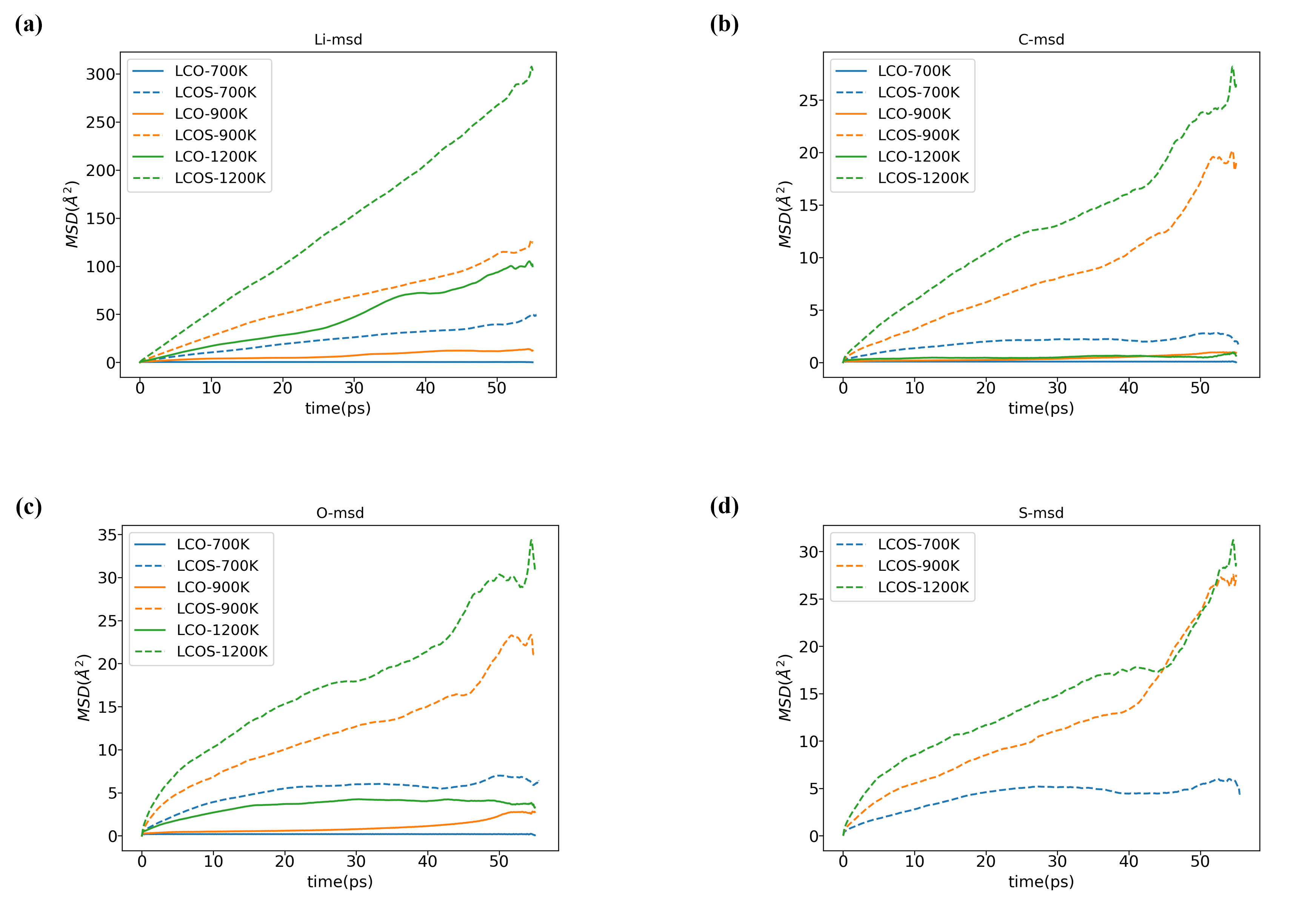}
    \caption{The MSD curves of (a) Li, (b) C, and (c) O in Li$_2$CO$_3$ and Li$_2$CO$_2$S, along with the MSD curves of (d) S in Li$_2$CO$_2$S at temperatures of 700~K, 900~K, and 1200~K.}
    \label{lcos-msd}
\end{figure*}

\noindent To compare the ionic transport properties in these two structures, we performed AIMD simulations for Li$_2$CO$_3$ and Li$_2$CO$_2$S supercells for 55~ps at 700~K, 900~K, and 1200~K, respectively. The MSD curves for Li, C, O, and S at different temperatures are shown in Fig.~\ref{lcos-msd}. It can be observed that Li$^+$ diffusion in Li$_2$CO$_2$S is significantly higher than in Li$_2$CO$_3$ at each temperature, leading to a higher ionic conductivity. However, Fig.~\ref{lcos-msd}b - Fig.~\ref{lcos-msd}d reveal the structural degradation of Li$_2$CO$_2$S at elevated temperatures, as characterized by pronounced migration of C, O, and S. In contrast, Li$_2$CO$_3$ maintains structural integrity across the investigated temperature range. Although both materials exhibit lower ionic conductivities compared to pristine LPSC, Li$_2$CO$_2$S shows moderate conductivity improvement to Li$_2$CO$_3$ attributed to Li–S interactions. Recent experimental studies\cite{zhang2023spontaneous,jiang2023enhancing} have demonstrated that gas-treated LPSC experiences a slight conductivity reduction. Jiang et al.\cite{jiang2023enhancing} reported decreases in ionic conductivity and electronic conductivity for CO$_2$-treated LPSC. Although the treatment induces minor conductivity losses, the significantly enhanced interfacial stability ultimately improves overall battery performance.

\section{Conclusions}
We investigated the gas-solid reaction mechanisms of LPSC solid-state electrolytes in CO$_2$ and mixed CO$_2$/O$_2$ atmospheres through combined AIMD simulations and MLFF-accelerated MD. Through these computational approaches, we determined the structural evolution of reaction products and evaluated their electronic and ionic transport properties. The results reveal different reaction pathways in varied atmospheres.  In pure CO$_2$ environment, the LPSC(100) surface predominantly forms Li$_2$CO$_2$S as the primary products, which is hard to evolve into carbonates. Under mixed CO$_2$/O$_2$ condition, O$_2$ adsorption constitutes the initial surface process, followed by subsequent reaction with adsorbed CO$_2$ molecules, generating carbonate -CO$_3$ units that organize into a Li$_2$CO$_3$-dominated coating layer. 
The two different types of coating layers exhibit vast differences in electronic and ionic conductivity, suggesting atmospheric control as an effective strategy for tailoring coating composition and functionality.  To generalize this gas-solid reaction strategy for broader solid electrolyte systems, we establish the following design criteria:(1) Stable adsorption configurations can form between the surface and gas molecules, for instance, in this work both CO$_2$ and O$_2$ spontaneously bonding with the LPSC(100) surface; (2) The generated products maintain ion migration capabilities to ensure the fast ion conduction of the whole system; (3) Low electronic conductivity of the products ensures effective blocking between the electrolyte and the electrode; (4) The electrochemical window stability of the generated products can be assessed based on the contacted cathode material. With these guidelines, the surface modification strategies through gas-solid reactions will be hopefully extended to diverse solid state electrolyte systems. Future work will focus on adjusting the interfacial compositions and specific configurations through the reaction path selection, and addressing critical properties of the interfacial products.

\section{Acknowledgment}
 
This work was supported by the Strategic Priority Research Program of Chinese Academy of Sciences 
(Grant No. XDB0500201), and by the National Natural Science Foundation of China (Grants Nos. 52172258,  
12134012, 12374067, and 12188101).  The numerical calculations in this study were partly carried out on the ORISE Supercomputer. The numerical calculations in this study were carried out on the SunRising-1 computing platform. 
\bibliography{sample.bib}

\end{document}


\title{Supplementary Information of ``Gas–solid Reaction Dynamics on Li$_6$PS$_5$Cl Surfaces: A Case Study of the Influence of CO$_2$ and CO$_2$/O$_2$ Atmospheres Using AIMD and MLFF Simulations''}
\author{Zicun Li}
\affiliation{College of Physics
Nanjing University of Aeronautics and Astronautics (NUAA)
Nanjing 211106, China}
\affiliation{Beijing National Laboratory for Condensed Matter Physics, Institute of Physics, Chinese Academy of Sciences, Beijing 100190, China}
\author{Xinguo Ren}\email{renxg@iphy.ac.cn}
\affiliation{Beijing National Laboratory for Condensed Matter Physics, Institute of Physics, Chinese Academy of Sciences, Beijing 100190, China}
\author{Jinbin Li}\email{jinbin@nuaa.edu.cn}
\affiliation{College of Physics
Nanjing University of Aeronautics and Astronautics (NUAA)
Nanjing 211106, China}
\author{Ruijuan Xiao}\email{rjxiao@iphy.ac.cn}
\affiliation{Beijing National Laboratory for Condensed Matter Physics, Institute of Physics, Chinese Academy of Sciences, Beijing 100190, China}
\author{Hong Li}
\affiliation{Beijing National Laboratory for Condensed Matter Physics, Institute of Physics, Chinese Academy of Sciences, Beijing 100190, China}

\maketitle
\onecolumngrid

\textbf{Adsorption energy of CO$_2$ on LPSC (100) surface}
\begin{figure}
        \includegraphics[width=1.0\columnwidth]{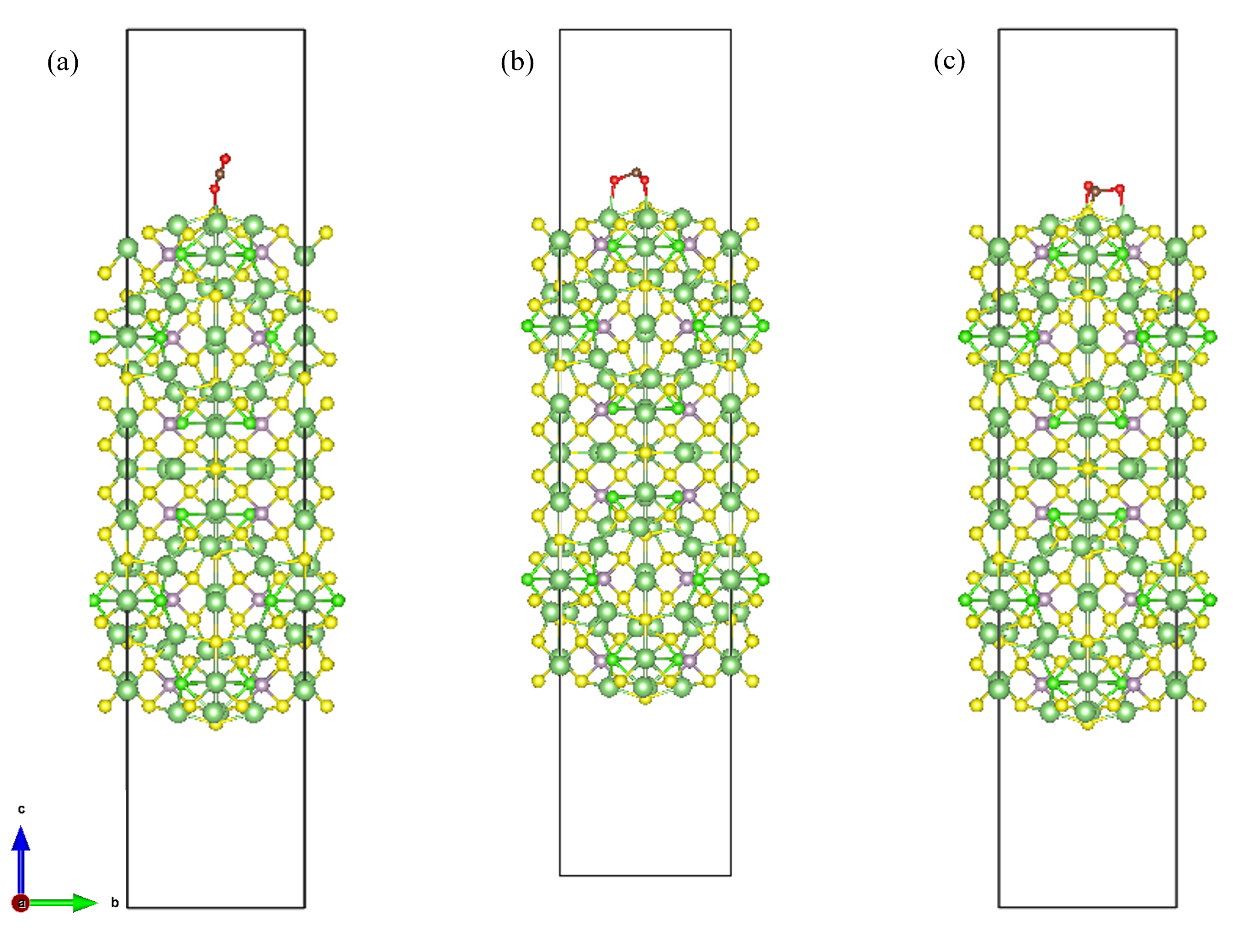}
        \caption{(a-c) is three types $CO_2$ adsorption sites on $Li_6PS_5Cl $(100) surface.}
        \label{fig:Sfig-100}  
\end{figure} 
\clearpage
\begin{figure}
    \centering
    \includegraphics[width=1.0\columnwidth]{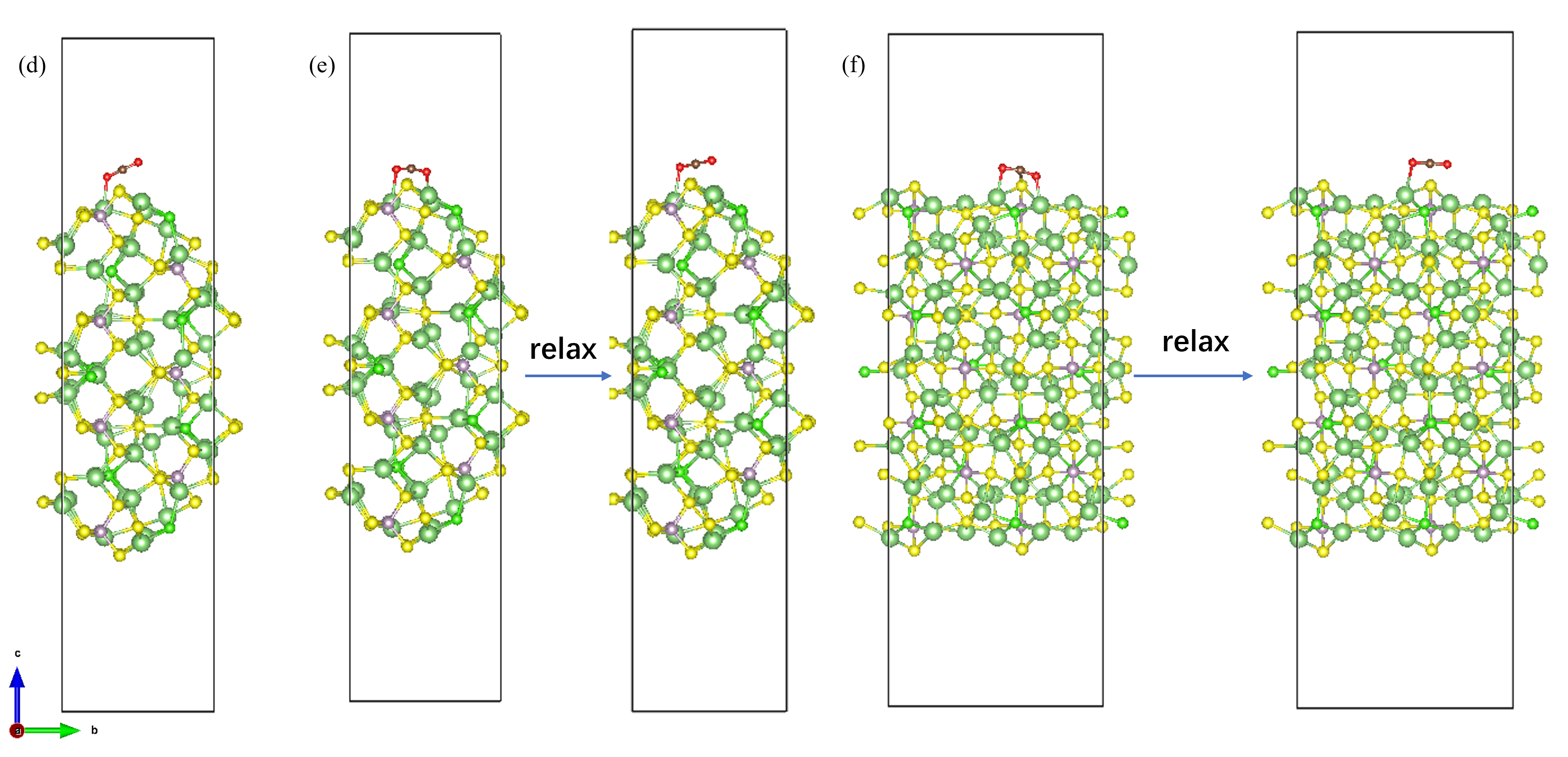}
    \caption{Three types adsorption sites of $CO_2$ on $Li_6PS_5Cl $(110) surface; and after structural relaxation of (e), (f) configuration.}
 \label{fig:Sfig-110}  
\end{figure} 

\begin{figure}
    \centering
    \includegraphics[width=1.0\columnwidth]{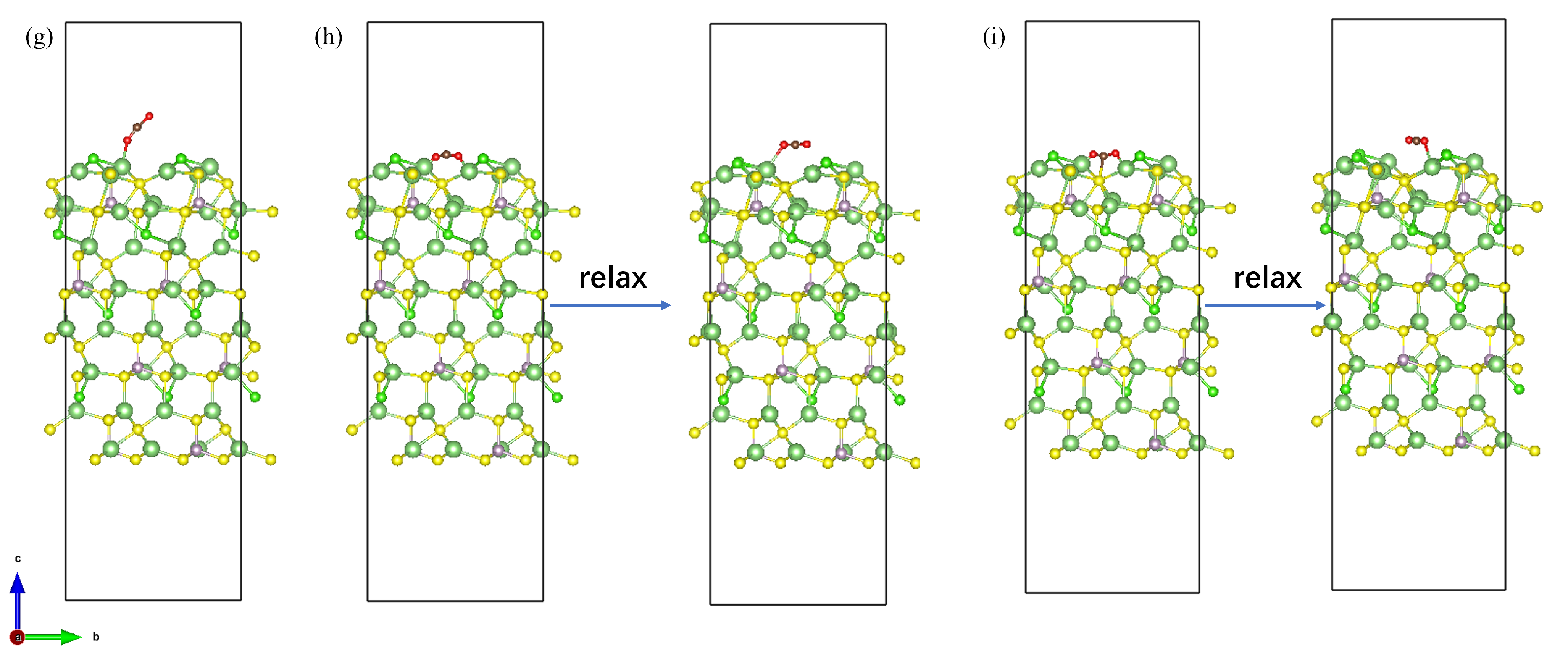}
    \caption{Three types adsorption sites of $CO_2$ on $Li_6PS_5Cl $(111) surface; and after structural relaxation of (h), (i) configuration. }
    \label{fig:Sfig-111}  
\end{figure} 
\clearpage

\begin{figure}
    \centering
    \includegraphics[width=1.0\columnwidth]{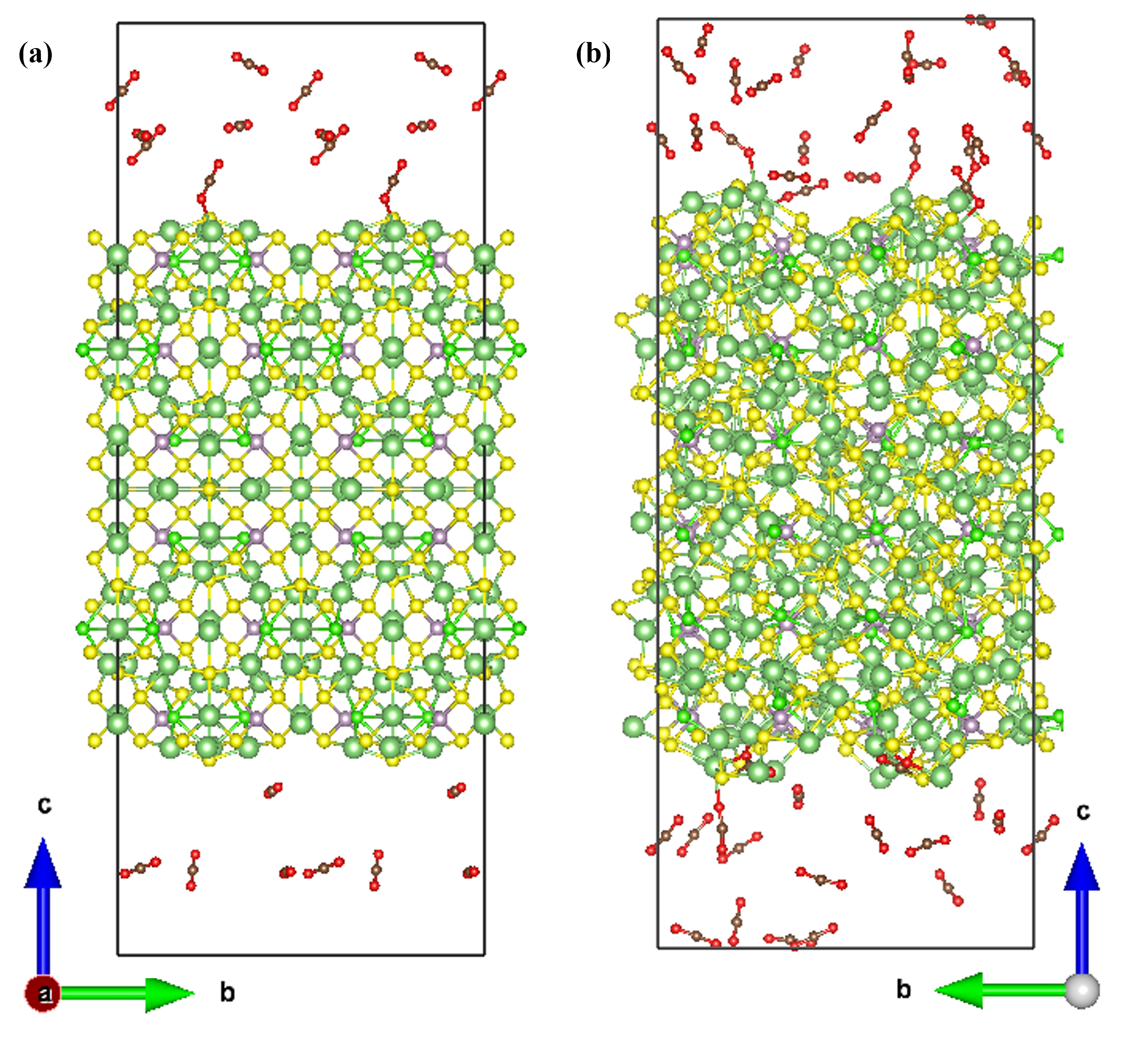}
    \caption{(a)The 2$\times$2$\times$1 supercell structure used for MLFF-based MD simulations; The structure after 100~ps MLFF-based MD simulations at (b)700~K.}
    \label{fig:lpsc-co2-222}  
\end{figure} 

\clearpage
\begin{figure}
    \centering
    \includegraphics[width=1.0\columnwidth]{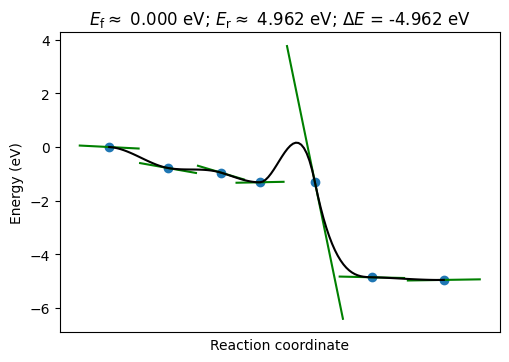}
    \caption{The energy of bond-breaking processes of $O_2$ adsorbed on the LPSC surface calculated by NEB.}
    \label{fig:O2_bond-broken}
\end{figure}
\clearpage

\begin{figure}
    \centering
    \includegraphics[width=1.0\columnwidth]{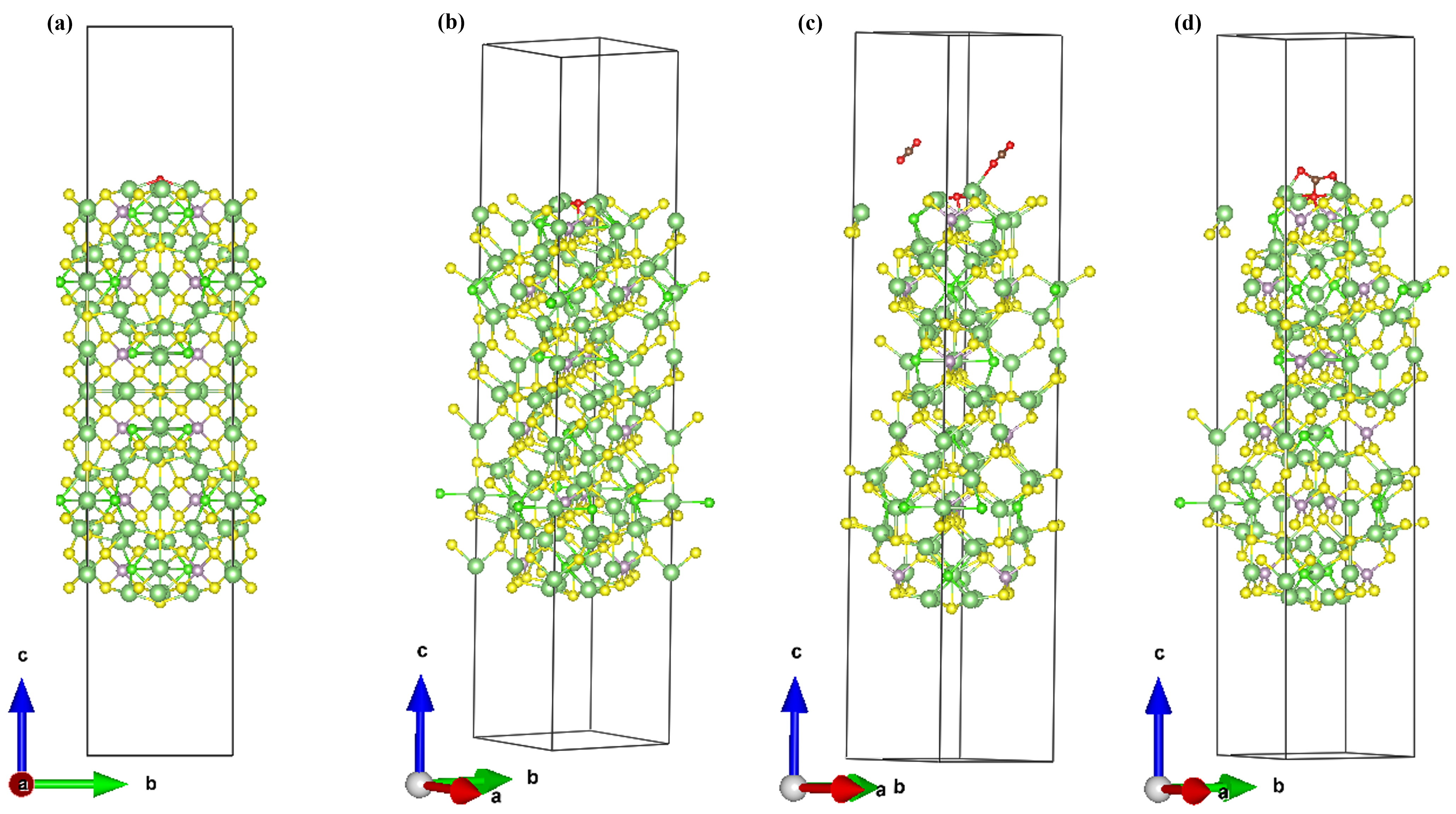}
    \caption{(a)The structure in which one O atom replace the free S atomic site; and perform structural relaxation on it(b); (c) one oxygen atom of CO$_2$ molecule closes to Li$^{+}$ on the relaxed structure; (d)two oxygen atoms are connected to two Li$^{+}$ ions with the bond formation between C atom of CO$_2$ and O atom on the surface.}
    \label{fig:LPSCO-ads}
\end{figure}

\textbf{Adsorption energy of CO$_2$ on LPSC(100) surface with O replace S}-The structure in which one O atom replace the free S atomic site is shown in Fig.~\ref{fig:LPSCO-ads} a. After structural relaxation of this configuration, the 
resulting structure is depicted in Fig.~\ref{fig:LPSCO-ads}b. Considering the two CO$_2$ adsorption modes on the LPSC surface- type\uppercase\expandafter{\romannumeral1} (Fig.~\ref{fig:LPSCO-ads}c) and type\uppercase\expandafter{\romannumeral3} (Fig.~\ref{fig:LPSCO-ads}d), the adsorption energies of CO$_2$ on the LPSCO surface are -0.10~eV and -2.48~eV, respectively.

\bibliography{sample.bib}